\documentclass{PoS}

\newcommand{\open}{\sphericalangle}

\title{Update on the phenomenology of collinear Dihadron FFs.}

\ShortTitle{DiFF Phenomenology}

\author{\speaker{A.~Courtoy}\\
        IFPA, AGO Department, Universit\'e de Li\`ege, B\^at. B5, Sart Tilman B-4000 Li\`ege, Belgium
\\  Divisi\'on  de Ciencias e Ingenier\'ias, Universidad de Guanajuato, C.P. 37150, Le\'on, M\'exico \\
        E-mail: \email{aurore.courtoy@ulg.ac.be}}

\abstract{We summarize recent results obtained thanks to the phenomenology of Dihadron Fragmentation Functions. The results include the update on the fitting techniques for both the Dihadron Fragmentation Functions themselves and the transversity PDF. The determination of the tensor charge of the nucleons as well as the impact of the latter on search for physics beyond the standard model are also discussed. So is the future extraction of the subleading-twist PDF $e(x)$ from JLab soon-to-be-released Beam Spin Asymmetry data. }

\FullConference{XXIII International Workshop on Deep-Inelastic Scattering,\\
		27 April - May 1 2015\\
		Dallas, Texas}

\begin{document}

Dihadron Fragmentation Functions (DiFF) encode information about the fragmentation process of a quark into a hadron pair (plus something else, undetected). Fragmentation has been hardly studied through models for pion pairs, the main knowledge about those DiFFs  being for now fits from data.
In particular, they can be extracted from the process
$e^+ e^- \to (\pi^+ \pi^-)_{\mbox{\tiny jet}} (\pi^+ \pi^-)_{\overline{\mbox{\tiny jet}}} X$~\cite{Courtoy:2012ry,Radici:2015mwa}, explored at Belle.

We define the total $P_h = P_1 + P_2$ and relative $R = (P_1-P_2)/2$ momenta of the pair, with $P_h^2 = M_h^2 \ll Q^2=-q^2 \geq 0$ and $q = k - k'$ the space-like momentum transferred and $z$ is the sum of the fractional energies carried by the two final hadrons.
The leading-twist cross section in collinear factorization, namely by integrating upon all transverse momenta but ${\bf R}_T$ and ${\bf \bar{R}}_T$, can be written as~\cite{Boer:2003ya}
\begin{equation}
\frac{d\sigma}{d\cos\theta_2 dz d\cos\theta dM_h d\phi_R d\bar{z} d\cos\bar{\theta} d\bar{M}_h d{\phi}_{\bar R}} = 
\frac{1}{4\pi^2}\, d\sigma^0 \, \bigg( 1+ \cos (\phi_R + \phi_{\bar{R}} ) \, A_{e^+e^-} \bigg) \; , 
\label{e:e+e-cross}
\end{equation}
where $\theta_2$ is the angle in the lepton plane formed by the positron direction and $\bf{P}_h$, and the azimuthal angles $\phi_R$ and $\phi_{\bar{R}}$ give the orientation of the planes containing the momenta of the pion pairs with respect to the lepton plane and the polar angle $\theta$ which is the angle between the direction of the back-to-back emission in the center-of-mass (cm) frame of the two final hadrons, and the direction of $P_h$ in the photon-proton cm frame. 
The $d\sigma^0$ is the unpolarized cross section producing an azimuthally flat distribution of pion pairs coming from the fragmentation of unpolarized quarks, $D_1$. In absence of data for the corresponding multiplicities, the unpolarized DiFFs have been parametrized to reproduce the two-pion yield of the 
{\tt PYTHIA} event generator tuned to the Belle kinematics~\cite{Courtoy:2012ry}. The term $A_{e^+e^-}$ represents the Artru-Collins asymmetry measured at Belle~\cite{Vossen:2011fk}. The latter is described in terms of chiral-odd fragmentation functions $H_1^{\sphericalangle}$, which were therefore extracted from Belle data for the first time three years ago~\cite{Courtoy:2012ry}.

\begin{figure}
\centering
\includegraphics[width=7cm]{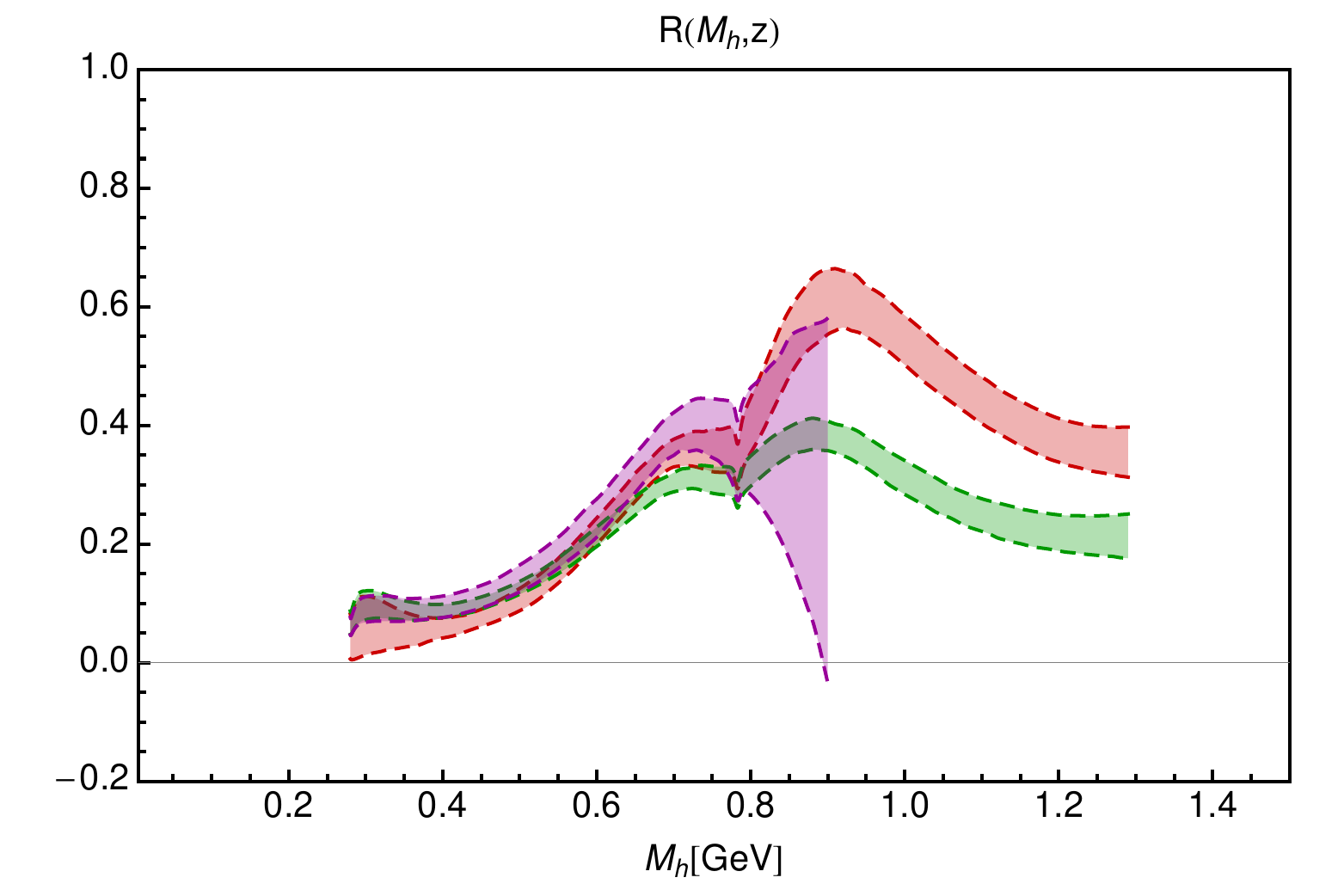} 
\caption{The ratio $R(z, M_h)$ as a function of $M_h$ at $Q_0^2=1$ GeV$^2$ for three different $z=0.25$ (shortest band), $z=0.45$ (lower band at $M_h \sim 1.2$ GeV), and $z=0.65$ (upper band at $M_h \sim 1.2$ GeV), with the value $\alpha_s (M_Z^2) = 0.139$ used in the QCD evolution equations. } 
\label{fig:H1Mh}
\end{figure}

An update of the fit of $H_1^{\sphericalangle}$ has been recently proposed in Ref.~\cite{Radici:2015mwa} in which
the error analysis is carried out using a Monte Carlo approach. This approach consists in creating $N=100$ replicas of the data points. In each replica, the data point in the bin $(z_i,M_{h\, j})$  is perturbated by a Gaussian noise with the same variance as the experimental measurement. Each replica, therefore, represents a possible outcome of an independent measurement.  The fit has been performed with the choice of hadronic scale $Q_0^2=1$GeV$^2$ and using two different values for $\alpha_s(M_Z^2)=0.125,\, 0.139$ in the QCD evolution routine. In Fig.~\ref{fig:H1Mh}, we show the new results for the ratio 
\begin{equation}
R(z, M_h) = \frac{|\bf{R}|}{M_h}\, \frac{H_1^{\open\, u} (z, M_h; Q_0^2)}{D_1^u (z, M_h; Q_0^2)} \; , 
\label{e:R}
\end{equation}
as a function of $M_h$ for chosen values of $z$ for $\alpha_s(M_Z^2)= 0.139$. The chisquare (per degree of freedom) distribution is centered around 1.6. The errorbands represent the $68$ central replicas. There is only a mild dependence on the value of the strong coupling. No significative difference with the standad Hessian method of our previous extraction is noticed.

\section{Transversity PDF}

\begin{figure}
\centering
\includegraphics[width=7cm]{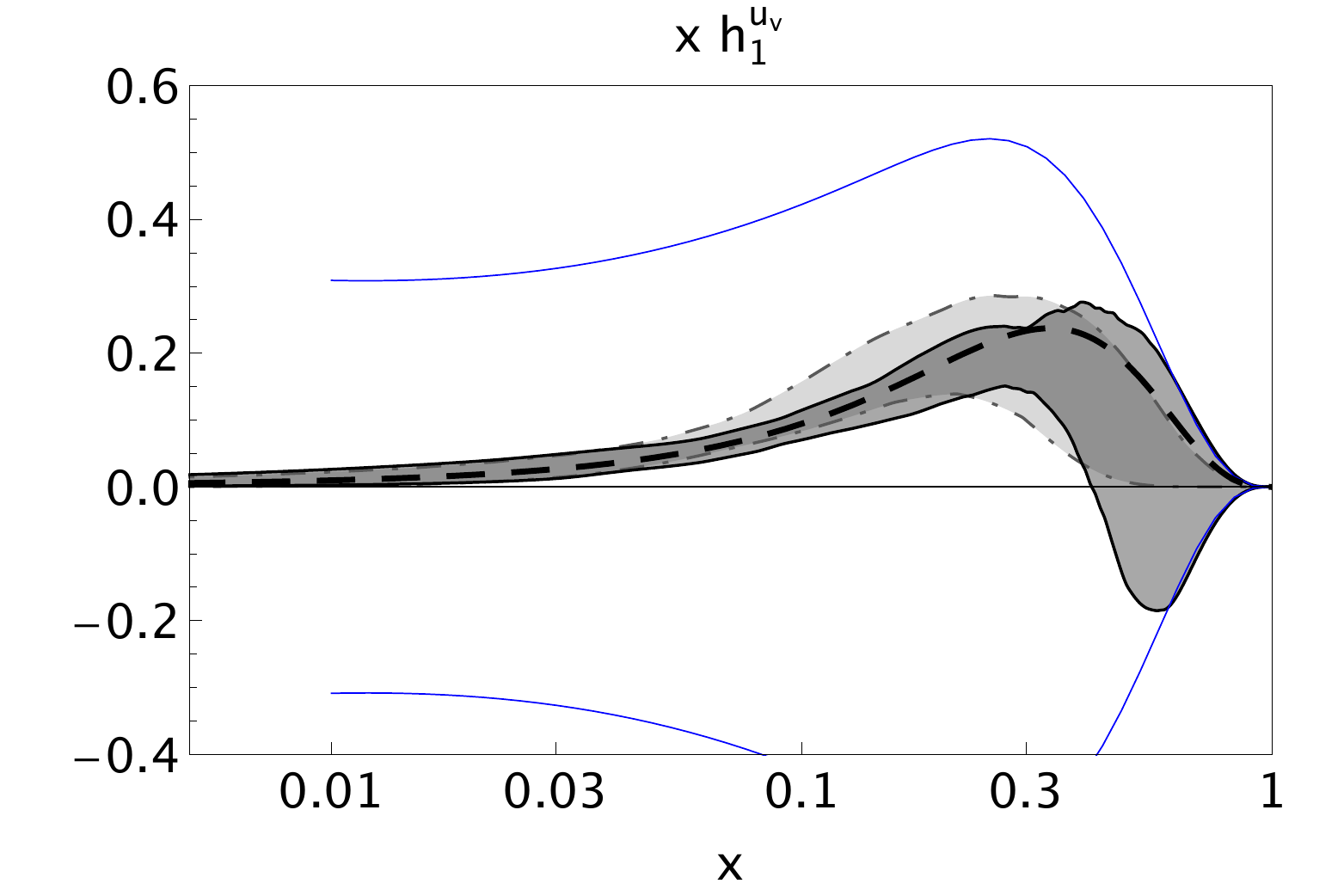} \hspace{0.5cm} \includegraphics[width=7cm]{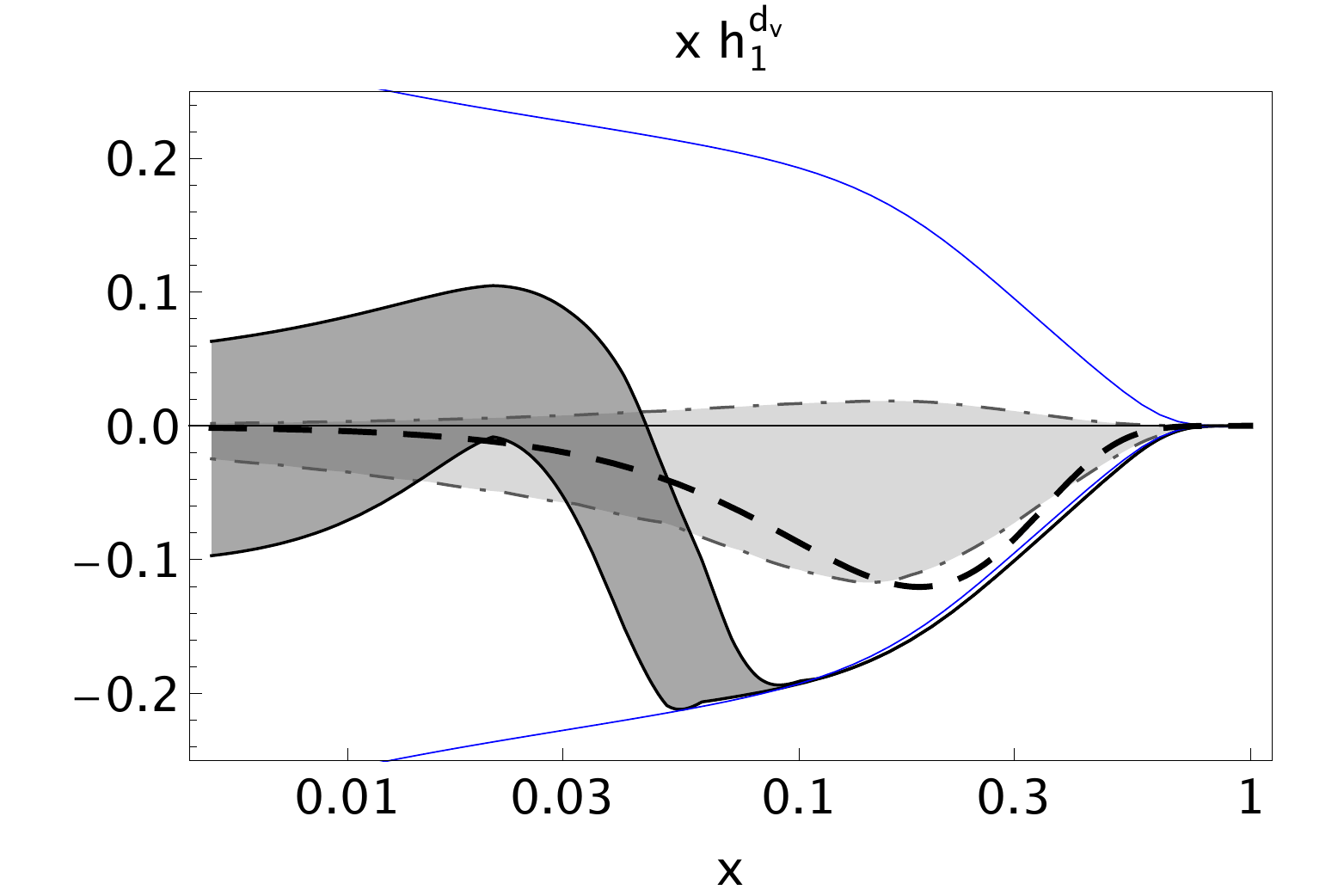}
\caption{The up (left) and down (right) valence transversities as functions of $x$ at $Q^2=2.4$ GeV$^2$. The darker band with solid borders in the foreground is our result in the flexible scenario with $\alpha_s (M_Z^2)=0.125$. The lighter band with dot-dashed borders in the background is the most recent transversity extraction from the Collins effect~\cite{Anselmino:2013vqa}. The central thick dashed line is the result of Ref.~\cite{Kang:2014zza}. The thick blue  lines indicate the Soffer bound.}
\label{fig:xhud}
\end{figure}

 Using $(\pi^+ \pi^-)$ SIDIS data off a transversely polarized proton target from HERMES~\cite{Airapetian:2008sk} and proton and deuteron targets from COMPASS~\cite{Adolph:2012nw} together with the results for the DiFFs, a point-by-point extraction of transversity was performed for the first time in the collinear framework~\cite{Bacchetta:2011ip} and an independent parameterization of the valence components of up and down quarks proposed in Ref.~\cite{Bacchetta:2012ty}. An update of the extraction of the valence transversities, using data for identified pions from COMPASS~\cite{Braun:2015baa}, was released together with the new parameterization of the DiFFs~\cite{Radici:2015mwa}. 
The process of interest is Dihadron SIDIS off transversly polarized targets,
\begin{equation}
 l + N \to l' + H_1 + H_2 + X\quad, \nonumber
 \end{equation}
where $l$ denotes the (unpolarized) lepton beam, $N$ the nucleon target, $H_1$ and $H_2$ the produced hadrons (here both are pions).  The data range in $x$ is restricted to $x\in [0.0065, 0.29]$. It allows one to access the (collinear) transversity, through  the modulation from the azimuthal angle $\phi_S$ of the target polarization  component $S_T$, transverse to both the virtual-photon and target momenta,
and $\phi_R$. The structure function attached to that modulation is written as the product of the PDF $h_1^q$, and the DiFF $H_1^{\sphericalangle\, q}$ 
~\cite{Bianconi:1999cd},
\begin{eqnarray}
 F_{UT}^{\sin (\phi_R +\phi_S)} (x, z, M_h; Q^2)& = & \, x\,
\sum_q e_q^2\,  h_1^q(x; Q^2)\,\frac{|\bf R| \sin \theta}{M_h} \, H_1^{\open\, q} (z, M_h; Q^2).
\end{eqnarray}  
Three different functional forms have been used for the parameterisation of the valence transversities, and their chisquare (per degree of freedom) distribution are centered in $[1.4, 2]$. The main output, {\it w.r.t} the previous extraction, is the improvement of the acurracy of the fit in the mid-range $x$ values for the up distribution, as reflected by the error band. The down quark valence distribution still tends to saturate the Soffer bound, a result that is somehow obtained in the single-pion SIDIS extraction with TMD evolution~\cite{Kang:2014zza} as well, as is illustrated in Fig.~\ref{fig:xhud}.

The tensor charge, defined as the first Mellin moment of the valence transversity PDF, is obtained by integrating over the physical support in $x$,
\begin{eqnarray}
\delta q_v (Q^2) &= &\int_0^1 dx \, h_1^{q_v} (x, Q^2) \; . 
\label{e:tensch}
\end{eqnarray}
 Notice that the error coming from the extrapolation of the PDF outside the data range is the main source of uncertainty as shown in Table 3 of Ref.~\cite{Radici:2015mwa}. The value for the isovector tensor charge $g_T = \delta u_v - \delta d_v$   for the functional form related to the so-called ``flexible scenario" with $\alpha_s (M_Z^2)=0.125$ is $g_T = 0.81 \pm 0.44$ at $Q^2=4$ GeV$^2$. It is in agreement with lattice determinations as well as with the other extractions from hadronic phenomenology \cite{Anselmino:2013vqa,Goldstein:2014aja,Kang:2015msa}, though the absolute value is slightly smaller than the lattice's.


\section{Impact of Tensor Charge on BSM}

The determination of the tensor charge has an impact on beyond the standard model (BSM) physics searches through high precision beta decay observables.
The latter allow us to probe couplings  other than of the $V-A$ type, which could appear at the low energy scale. 
Experiments using cold and ultra-cold neutrons, nuclei,  and meson rare decays~\cite{Baessler:2014gha}, are being performed, or have been planned,  that can reach the per-mil level or even higher precision.
These low energy measurements  can be connected to BSM effects generated at TeV scales through effective field theories~\cite{Bhattacharya:2011qm}. In this approach that  complements collider searches,
the new  interactions are introduced in an effective Lagrangian describing semi-leptonic transitions at the GeV scale including  four-fermion terms,
or operators up to dimension six  for the scalar, tensor, pseudo-scalar, and V+A interactions, with  effective coupling $\epsilon_i$,  ($i=S,T,P,L,R$) ,  {\it e.g.}~\cite{Cirigliano:2013xha}.

\begin{center}
\begin{figure}
\centering
\includegraphics[width=7.cm]{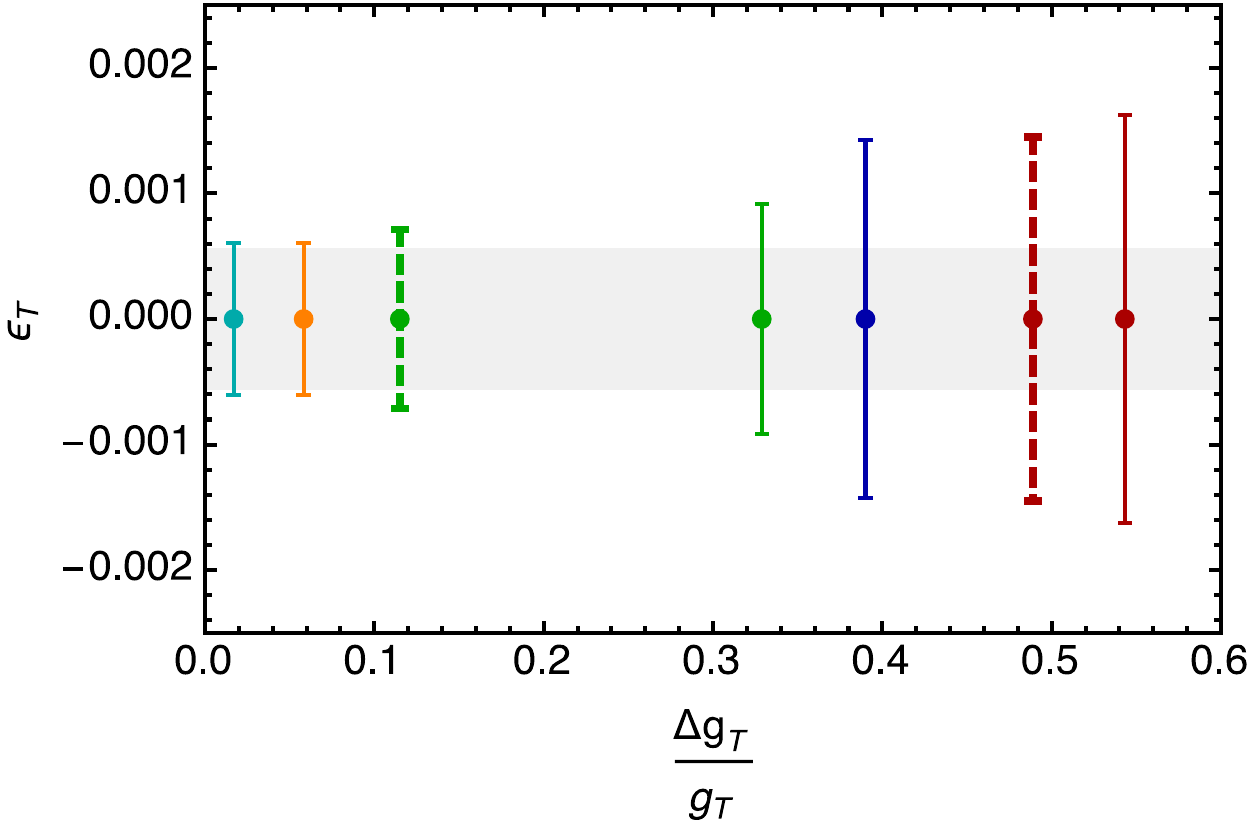}
\caption{Bounds on $\epsilon_T$ using various values of $g_T$: (turquoise-yellow)  Lattice QCD; (green) Deeply virtual $\pi^0$ and $\eta$ production \cite{Goldstein:2014aja}; (blue) single pion SIDIS \cite{Anselmino:2013vqa};  (red) dihadron SIDIS \cite{Radici:2015mwa}. 
The dashed lines are for the future projections of the experimental extractions based on projections for JLab@12.
The grey band gives the uncertainty in $\epsilon_T$ assuming $\Delta g_T=0$ and the (weighted) average value for $g_T$ using the three lattice results plus the three hadronic results. }
\label{fig:et}
\end{figure}
\end{center}

The scalar (S) and tensor (T) operators  contribute linearly  to the beta decay parameters through their interference with the SM amplitude, and they are therefore more easily detectable. 
The matrix elements/transition amplitudes  between neutron and proton states of all quark bilinear Lorentz structures in the effective Lagrangian which are relevant for  beta decay observables, 
involve products of the BSM  couplings, $\epsilon_i$,  and the corresponding hadronic charges, $g_i$, {\it i.e.} considering  only terms with left-handed neutrinos
\begin{eqnarray}
\label{eq:JTW}
\Delta{\cal L}_{eff} = 
&&\!\!\!\!\!\!-\, C_S~ \bar{p} n \cdot \bar{e} (1 - \gamma_5) \nu_e -\, C_T~ \bar{p}\sigma_{\mu\nu} n \cdot \bar{e}\sigma^{\mu\nu} (1 - \gamma_5) \nu_e~,
\end{eqnarray}
where $C_S =   G_F V_{ud} \sqrt{2} \epsilon_{S}  g_{S}$, and $C_T =   4 G_F V_{ud} \sqrt{2} \epsilon_{T}  g_{T}$\footnote{The latest global bounds on $C_T$ are  given in Refs.~\cite{Pattie:2013gka}.}. The tensor charge as determined here through dihadron SIDIS can be used to study the error on the BSM tensor effective coupling, currently still compatible with zero. In Ref.~\cite{Courtoy:2015haa}, the three hadronic extractions of the tensor charge ---{\it i.e.} single-hadron SIDIS, dihadron SIDIS, $\pi^0$ and $\eta$ in DVMP--- have been considered to set an experimentally based limit on the possible new interaction, given the experimental bounds on $C_T$. Though the lattice simulations still give, in that particular case, the most strigent bounds, we have shown, see Fig.~\ref{fig:et}, that hadronic phenomenology will be competitive thanks to future experiments. 

As for future extractions, the dihadron SIDIS  will be studied in CLAS12 at JLab on a proton target and in SoLID on a neutron target that will give together an improvement  of $\sim 10\%$ in the ratio $\Delta g_T/g_T$ thanks to a wider kinematical coverage and better measurement of the $d$ quarks contribution~\cite{longversion}. This estimation is based on a reanalysis of the fitting output with pseudo-data generated from projections with stastical and systematic errors for the mentionned experiments\footnote{See JLab proposals.}. While the absolute value of the transversity can obviously not be determined yet, the magnitude of the predicted error is what matters here. 
%
%
%
%

\section{Twist-3 PDF}

DiFFs also appear in asymmetries involving subleading objects. An extraction of the twist-3 PDF, $e(x)$, through the  analysis of the  preliminary data~\cite{Pisano:2014ila} for the $\sin\phi_R$-moment of the beam-spin asymmetry for  dihadron Semi-Inclusive DIS at CLAS at 6 GeV was proposed in Ref.~\cite{Courtoy:2014ixa}. Pion-pair production off unpolarized target in the DIS regime provide an access to the higher-twist Parton Distribution Functions $e(x)$ and to  Dihadron Fragmentation Functions. 

The  $e(x)$ PDF offers important insights into the physics of the largely-unexplored quark-gluon correlations, and its $x$-integral is related to the marginally-known
scalar-charge of the nucleon, and to the pion-nucleon $\sigma$-term. 

\begin{figure}[t]
\begin{center}
		\includegraphics[scale= .5]{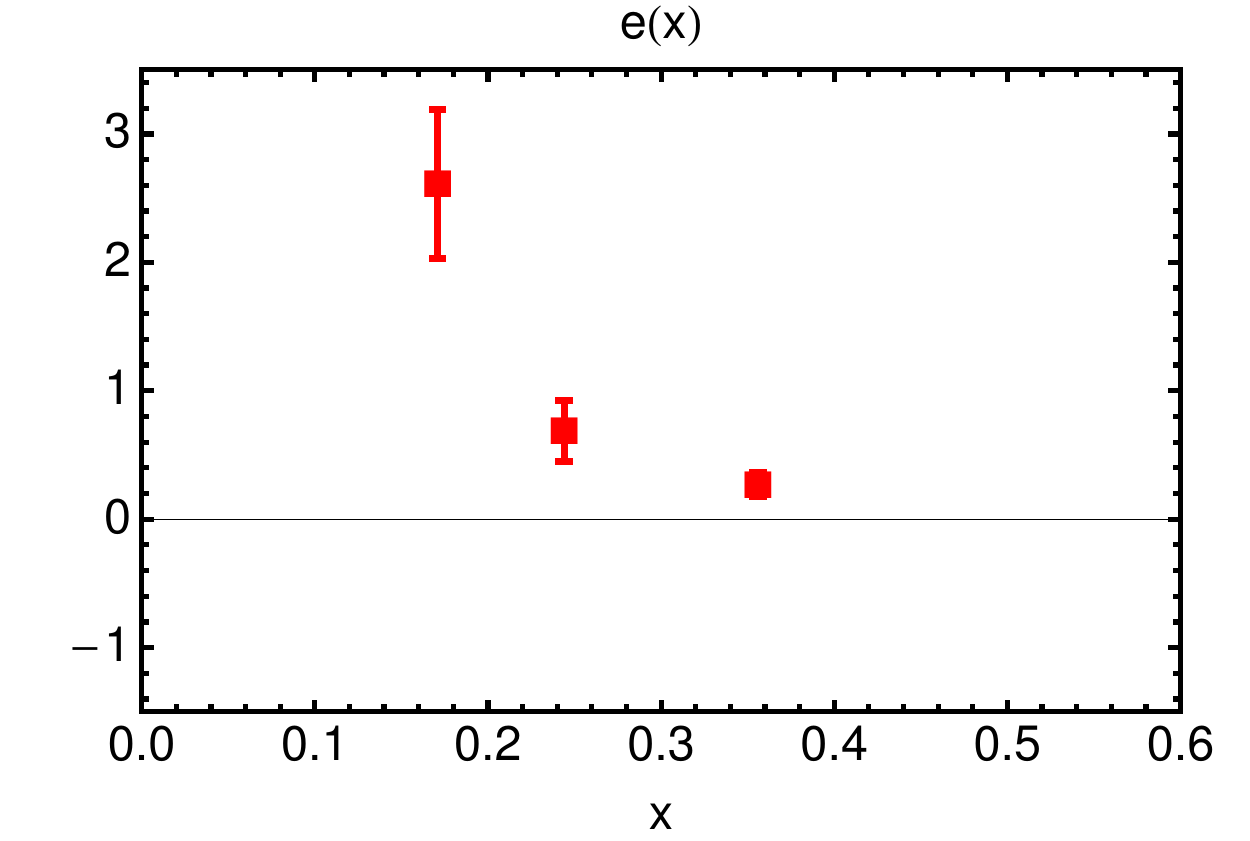}
\end{center}
\caption{  The extraction of the combination $e^V\equiv4 e^{u_V}(x_i, Q_i^2)/9-e^{d_V}(x_i, Q_i^2)/9$ in the WW scenario. The error bars correspond to the propagation of the experimental and DiFF errors. }
\label{fig:e_extract_us_ww}
\end{figure}

Beam Spin Asymmetry $A_{LU}$ presented in the proceedings~\cite{Pisano:2014ila},
to leading-order in $\alpha_s$ and leading term in the Partial Wave Analysis, is expressed in terms of FFs and collinear PDFs~\cite{Bacchetta:2003vn}
\begin{eqnarray}
&&A_{LU}^{\sin \phi_R } \left( x,  z, M_{h} ; Q^2, y \right) \nonumber\\
&=&-\frac{W(y)}{A(y)}\,\frac{M}{Q}\,\frac{|\bf{R}  |}{M_{h}} \, 
\frac{ \sum_q\, e_q^2\, \left[ x e^q(x; Q^2)\, H_{1}^{\sphericalangle, q}(z, M_{h}; Q^2)  + \frac{M_{h}}{z M} \,  f_1^q(x; Q^2)\, 
          \tilde{G}^{\sphericalangle, q}(z, M_{h}; Q^2) \right]  } 
       { \sum_q\, e_q^2\,f_1^q(x; Q^2)\, D_{1}^q (z, M_{h}; Q^2) }\quad ,\nonumber\\
 \label{eq:alu}
\end{eqnarray}
The dependence in $(z, M_{h})$ is factorized in the DiFFs and kinematical factors, leaving the dependence in $x$ for the PDFs.
The twist-2 functions are $f_1(x), H_1^{\open} (z, M_{h})$ and $D_1(z, M_{h})$, while the twist-3 functions are $e(x)$ and $\tilde{G}^{\sphericalangle}(z, M_h)$. Under the assumption, related to the Wandzura-Wilczek approximation, that the the twist-3 DiFFs are negligible, the PDF $e(x)$ is directly extracted, point-by-point. The results are shown on Fig.~\ref{fig:e_extract_us_ww}. The average scale of the data is $Q^2\sim 1.5$ GeV$^2$. A recent calculation in the LFQM~\cite{Lorce:2014hxa}, in which only the kinematical part of the twist-3 PDF contributes, shows a good agreement with the present extraction, when evolved to $1.5$ GeV$^2$. The official publication of the data as well as complementary and future experiments will shed more light on the twist-3 sector of the DiFF phenomenology.

\section{Conclusions}

We have presented various aspects of the rich phenomenology involving dihadron fragmentation functions. Though dihadron SIDIS represent a (consequently smaller) subset of single-hadron SIDIS, the results achieved through these processes have been very promising so far. Future analyses ---{\it e.g.} dihadron multiplicities at COMPASS--- and experiments ---especially in CLAS12 and SoLID at JLab--- will increase the data set, allowing for an improved knowledge on both DiFFs at leading and subleading-twist and collinear leading and subleading PDFs. 

\acknowledgments

The author acknowledges her  co-authors in Pavia, A. Bacchetta and M.~Radici, discussions and broad collaboration with S.~Liuti and S.~Pisano. She thanks her co-authors S.~Baessler and M.~Gonz\'alez-Alonso for interesting discussions. A.C. would like to thank the collaborations with SoLID and CLAS12 in general.

\end{document}